# On the role of boron, carbon and zirconium on hot cracking and creep resistance of additively manufactured polycrystalline superalloys

Arthur Després[a*], Stoichko Antonov[b], Charlotte Mayer[c], Catherine Tassin[a], Jean-Jacques Blandin[a], Paraskevas Kontis[b], Guilhem Martin[a]

[a]Univ. Grenoble Alpes, CNRS, Grenoble INP, SIMaP, F-38000 Grenoble, France
[b]Max-Planck-Institut für Eisenforschung GmbH, Max-Planck-Strasse 1, 40237 Düsseldorf, Germany
[c]Aubert et Duval, Usine des Ancizes, Rue des villas, BP1, 63770 Les Ancizes, France

**Abstract**

We investigate the hot cracking susceptibility and creep resistance of three versions of a nickel-based superalloy with different contents of boron, carbon and zirconium fabricated by laser powder bed fusion. Crack-free and creep resistant components are achieved for alloys with boron, carbon and no zirconium. We then rationalize this result by evaluating how boron, carbon and zirconium are distributed at grain boundaries in the as-built and heat-treated microstructures of an alloy containing all these elements. Observations are conducted by scanning and transmission electron microscopy, and atom probe tomography. In the as-built microstructure, boron, carbon and zirconium segregate at high-angle grain boundaries as a result of solute partitioning to the liquid and limited solid-state diffusion during solidification and cooling. After heat-treatment, the amount of boron and carbon segregating at grain boundaries increases significantly. In contrast, zirconium is not found at grain boundaries but it partitions at the γ' precipitates formed during the heat treatment. The presence of zirconium at grain boundaries in the as-built condition is known to be susceptible to enhance hot cracking, while its absence in the heat-treated microstructure strongly suggests that this element has no major effect on the creep resistance. Based on our observations, we propose alloy design guidelines to at the same time avoid hot cracking during fabrication and achieve the required creep performance after heat-treatment.

**Keywords:** additive manufacturing; Ni-based superalloys; hot cracking; creep; grain boundary segregation

―――――――

[*] Corresponding author, *E-mail address:* arthur.despres@grenoble-inp.fr.

# 1. Introduction

Additive manufacturing of nickel-based superalloys currently stimulates significant efforts to develop new compositions adapted to this kind of processing route [1–5]. The fast cooling rates and strong thermal constraints taking place during processing renders many polycrystalline γ'-strengthened nickel-based superalloys sensitive to hot cracking, i.e. the formation of cracks in the bulk of the microstructure during fabrication. This phenomenon arises regardless of the fabrication technique employed, see e.g. electron powder bed fusion [6], laser powder bed fusion [1,5,7,8] and direct metal deposition [9,10]. This poses a major challenge for the alloy design because strategies to avoid hot cracking generally lead to an unacceptable decrease of the creep resistance.

Hot cracks develop during the last stages of solidification when the thermal constraints exert tension forces on the residual liquid films located at grain boundaries, see e.g. [11,12]. This issue is common to all kinds of processes involving solidification, and much of the current knowledge is derived from directional solidification casting and welding experiments. The hot cracking susceptibility of nickel-based superalloys is for example known to be very sensitive to phosphorous, sulphur, boron, carbon and zirconium [13–15], i.e. intentionally and unintentionally added trace elements that have a low solubility in the fcc-γ matrix and $L1_2$- γ'-precipitates [16].

The partitioning of these elements in the liquid during solidification lowers the end-of-solidification temperature and increases the chance of hot crack initiation. Historically, phosphorous and sulphur were considered as the major promoters of hot cracking during solidification of welds [15], but the content of these elements has been reduced to a negligible fraction in modern alloys thanks to progresses in refining metallurgy. Nowadays, the attention is increasingly focused on the role of boron, carbon and zirconium. According to several castability tests, boron and zirconium strongly promote hot cracking, in particular when both are present in the nominal composition [13,14]. The role of carbon is still debated, and it has been proposed that it might aid to avoid hot cracking, see e.g. [13,17]. In additive manufacturing applications, two main strategies have been implemented to control the effect of minor elements on hot cracking. The first strategy is to optimize their content in the bulk alloy composition [1,13]. The second strategy consists in altering the solidification conditions to increase the grain boundary density. This allows grain boundary solutes to be distributed over more interfaces and thus maintain grain boundary segregation below a critical level promoting hot cracking [18].

One has to remain careful in implementing these strategies because the main purpose of adding minor elements such as boron, carbon and zirconium to nickel-based superalloys is to improve the creep lifetime of the component after heat-treatment. This aspect is paramount for polycrystalline superalloys, where the time to rupture may vary by one or two orders of magnitude as a function of the exact content of these elements [19–21]. The mechanisms causing the extreme sensitivity of creep resistance to these elements are still debated, but it is generally admitted that they relate to the strengthening of grain boundaries. Under thermodynamic equilibrium conditions, boron segregates at grain boundaries in the form of solid solution [22,23] and can also promote the formation of borides enriched in chromium, molybdenum and tungsten [21,24]. The presence of boron at grain boundaries is required to optimize creep properties, while the role of borides remains unclear since a too large increase in the boron content promotes boride precipitation but also deteriorates the creep resistance [21]. Carbon has a similar behaviour as it can be found in solid solution at grain boundaries [25], or in various carbides and carbonitrides [19]. Carbide or carbonitride precipitation raises the material's strength, and may induce grain boundary serrations, which is beneficial for creep properties [26]. The role of zirconium is less well known, and has been related either to grain boundary segregation [19], or the formation of compounds with elements like sulphur and oxygen which would otherwise deteriorate the mechanical properties [27]. Overall, it is not possible to achieve the creep resistance required for polycrystalline nickel-based superalloys without adding a small fraction of these elements to the bulk composition.

Current studies on polycrystalline nickel-based superalloys produced by additive manufacturing focus mainly on understanding the role of grain boundary segregation on the hot cracking susceptibility during fabrication [1,3,6,18]. However, very little attention has been given to the evolution of grain boundary segregation from the as-built to the heat-treated state and its effect on creep resistance. As the microstructure after fabrication is strongly out-of-equilibrium, an extensive redistribution of solutes during the heat-treatment is expected. The objective of the present study is to investigate this redistribution of solutes and to link it to the hot cracking susceptibility and creep resistance.

In a first step, we examined the hot-cracking susceptibility and creep lifetime of three versions of a hard to weld polycrystalline superalloy. In these three versions, the content in boron, carbon and zirconium has been intentionally modified. The samples were fabricated by laser powder bed fusion, a process leading to fast cooling rates (in the range of $10^5$ K.s$^{-1}$ to $10^7$ K.s$^{-1}$ [5,28]) and limited in-situ heat-treatment. In a second step, we investigated for one version of this superalloy how boron, carbon and zirconium are distributed in the as-built and heat-treated conditions. The microstructural observations are followed by a discussion on the origin of these solute distributions and on their influence on hot cracking and creep resistance. From these results, guidelines to design crack-free and creep resistant polycrystalline nickel-based superalloys for additive manufacturing are suggested.

## 2. Experimental procedures

*2.1. Powder composition and process parameters*

Three powder batches of the γ'-hardened AD730® alloy were gas-atomized and supplied by Aubert et Duval. Their nominal composition is given in Table 1. Carbon, nitrogen and oxygen were measured by thermal conductivity-infrared absorption, while the other elements were measured by atomic emission spectroscopy using inductively coupled plasma. About 0.010 wt.% of oxygen and 0.005 wt.% of nitrogen have also been detected, as a result of surface adsorption during powder fabrication. The sulphur content was kept below 10 ppm for all three powder batches.

The first version of this alloy is labelled (B+C+Zr) as it contains boron, carbon and zirconium in significant fractions and exhibits a chemical composition within the standard range of the AD730® alloy processed by the traditional cast and wrought route [29]. The second version is labelled (B+C) since its zirconium content has been reduced to negligible levels. The version (C) contains negligible levels of boron and zirconium. Note that the content in these trace elements is not constant between two different versions of the alloy, and that the content in major elements also varies within 0.3-0.5 wt.% between the three versions of the alloy. Given the known extreme sensitivity of hot cracking susceptibility and creep resistance to the minor elements [19–21], the variations in major elements are unlikely to explain the drastic changes in hot-cracking and creep resistances presented below. However, this aspect will be considered during the discussion of the results.

Table 1: Composition of the three alloy versions. The three first rows of composition are in wt.%, while the last three rows are in at.%. '-' denotes a fraction below 0.001%.

| Version | Ni | Al | Ti | Cr | Fe | Co | Nb | Mo | W | B | C | Zr |
|---|---|---|---|---|---|---|---|---|---|---|---|---|
| (B+C+Zr) | bal. | 2.3 | 3.1 | 15.9 | 4.1 | 8.5 | 1.0 | 3.0 | 2.7 | 0.009 | 0.022 | 0.045 |
| (B+C) | bal. | 2.3 | 3.1 | 16.5 | 4.1 | 8.5 | 0.5 | 3.1 | 3.1 | 0.004 | 0.038 | - |
| (C) | bal. | 2.2 | 3.6 | 16.2 | 3.6 | 8.1 | 1.0 | 3.0 | 3.1 | - | 0.010 | - |
| (B+C+Zr) | bal. | 5.0 | 3.7 | 17.5 | 4.2 | 8.2 | 0.6 | 1.8 | 0.8 | 0.050 | 0.104 | 0.028 |
| (B+C) | bal. | 4.9 | 3.7 | 18.1 | 4.2 | 8.3 | 0.3 | 1.8 | 1.0 | 0.023 | 0.181 | - |
| (C) | bal. | 4.9 | 4.3 | 17.9 | 3.8 | 7.9 | 0.6 | 1.8 | 1.0 | - | 0.050 | - |

## 2.2. Fabrication and heat treatment

Samples of 13.0mm in width, 10.0mm in height and 100.0mm in length were fabricated by laser powder bed fusion using an EOS M290 machine. The laser scan direction was rotated by 67° between each deposited layer. The laser power and scan speed were chosen after a parametric study aiming to minimize porosity, and are in the range of those reported by previous work on polycrystalline superalloys [1,2,7]. The final porosity density measured by image analysis was found to be about $0.03 \pm 0.02$ % $mm^{-2}$. All samples examined in the present work had a similar density. The fabrication parameters were kept identical for the three alloy versions.

After fabrication, the samples were subjected to a sub-solvus heat-treatment at 1080°C for 4 h, followed by ageing at 760°C for 16 h. This cycle of heat-treatment is standardly applied to the cast & wrought AD730® alloy, see [29]. It has been selected to avoid recrystallization, hence the solute redistribution is not impacted by a major evolution of the grain boundary density. It should be noted that more refined heat-treatment cycles could be implemented if one seeks to optimize the mechanical properties of components specifically produced by laser powder bed fusion.

### 2.3. Creep tests at 650°C

Cylindrical creep samples with reduced sections of 22.5 mm in length and 4.5 mm in diameter were prepared from the heat-treated samples, with the loading direction perpendicular to the build direction. The tests were carried out at 650°C under a load of 690 MPa, following the NF EN 2002-05 standard. Two samples were tested for each alloy version. The creep performance was evaluated based on the time to rupture $t_r$ and the Larson-Miller parameter $(\log(t_r) + 20) \times T)/1000$ This parameter provides a measure of the creep resistance normalized by the testing temperature $T$, and is frequently used for nickel-based superalloys [2,30].

### 2.4. Optical microscopy

Crack densities in the as-built and heat-treated microstructures were measured by image analysis using the ImageJ software from optical micrographs with large fields of view acquired using an Olympus DSX510 microscope. Specimens were ground and polished using abrasive media with a surface finish using 1µm alumina particles. A binary image was achieved using a Maximum Entropy automatic threshold. Objects with a circularity below 0.5 were considered as cracks. The crack density of a given sample was obtained by dividing the sum of all cracks Feret length by the total area investigated.

### 2.5. Scanning and transmission electron microscopy

Microstructural observations of the as-built and heat-treated specimens were conducted using a Zeiss Gemini SEM500 scanning electron microscope (SEM). Specimens were ground and polished using abrasive media to a 0.03 µm colloidal silica surface finish. Images were formed using the secondary electron (SE) and backscattered electron (BSE) modes. Inverse pole figure (IPF) maps were obtained from electron backscattered diffraction (EBSD) operating at 20 kV with a step size of 0.5 µm. Grains were identified as regions of the microstructure surrounded by boundaries of misorientation angle higher than 15°. The grain size was evaluated from the arithmetic mean of the grains equivalent area diameter. On the same samples, chemical analysis at the grain scale were performed using dispersive X-ray spectrometry (EDX) setting the beam acceleration voltage at 10 kV.

Thin foils of material were prepared in a plane containing the build direction, and observed in the TEM, on a JEOL 2100F at 200 kV accelerating voltage.

*2.6. Atom probe tomography*

The segregation of solutes at grain boundaries was investigated by atom probe tomography (APT). Site specific lift-outs were performed using an FEI dual beam focused ion beam Helios 600i. The APT specimens were extracted from the uncracked region ahead of a partially cracked high angle grain boundary in both the as-built and heat-treated samples. The specimens were prepared from a high angle grain boundary of 41° misorientation angle for the as-built sample, and a 45° misorientation angle for the heat-treated sample. The APT specimens were analyzed using a Cameca LEAP 5000 XR instrument operating in a laser mode with pulse rate at 125 kHz, pulse energy 45-60 pJ and base temperature 60 K. The commercial package AP Suite 6.0 was used for data reconstruction and analyses.

**3. Experimental Results**

*3.1. Hot cracking in the as-built state*

Figure 1 shows micrographs of the as-built microstructures of the three alloys investigated. Optical micrographs with large fields of view are also provided in supplementary materials (Figure S. 1) to give a more representative picture of the density of cracks in the three alloys. The (B+C+Zr) alloy exhibits hot cracks (Figure 1a), while the other two alloys are uncracked (Figure 1b,c). The dendritic morphology of the crack surface, in the enlarged view of Figure 1a, is typical of hot cracking in nickel-based superalloys processed by laser powder bed fusion [1,2,31].

The crack density in the as-built microstructure of the (B+C+Zr) alloy has been estimated from the optical micrographs to be $0.124 \pm 0.03$ mm/mm$^2$, while that in the heat-treated was measured to be $0.123 \pm 0.03$ mm/mm$^2$ (see Table S. 1 in the supplementary materials). These values correspond to a rather low crack density compared to previous reports [1,7]. By contrast, the as-built and heat-treated microstructures of the two other alloys do not contain cracks. The stability of the crack density with heat-treatment for the (B+C+Zr) alloy and the absence of cracks in the heat-treated microstructures of the two other alloys indicate that the AD730® alloy is not sensitive to strain age-cracking.

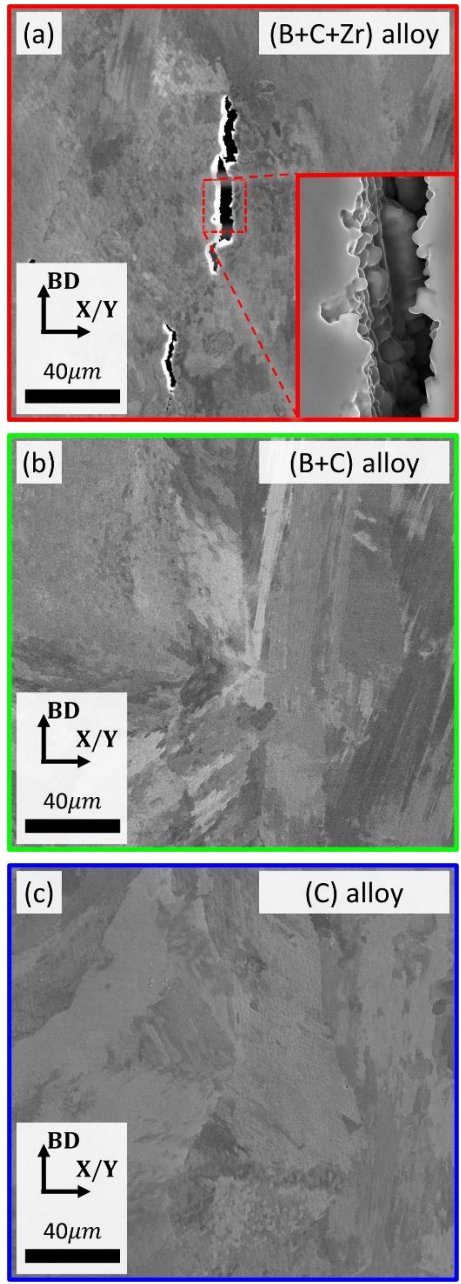

Figure 1: SEM-SE contrast micrographs of the as-built microstructure of (a) the (B+C+Zr) alloy, (b) the (B+C) alloy, (c) the (C) alloy. An enlarged view of the crack surface is shown in (a). The enlarged view was acquired with an in-lens detector.

*3.2. Creep resistance in the heat-treated state*

Figure 2 shows the creep resistance of the samples after heat-treatment plotted using their Larson-Miller parameter. The times to rupture of the different tests are also summarized in Table 2. The (B+C) alloy exhibits the best creep resistance, with times to rupture above 2000h, while the (B+C+Zr) alloy exhibits a moderate creep life, and the (C) alloy has an almost null resistance to creep although it was found to be free of cracks. The grain size of the heat-treated samples was calculated from EBSD maps to be $20 \pm 1.3$µm in equivalent diameter between the three versions of the investigated alloy. The large scattering of the times to rupture in the (B+C+Zr) alloy is possibly related to the presence of hot-cracks in the gauge length of the sample prior to the creep test.

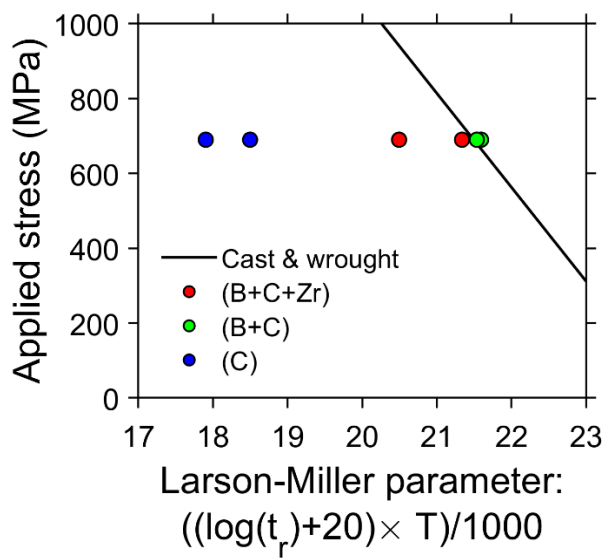

Figure 2: Creep resistance of the heat-treated samples plotted in terms of Larson-Miller parameter. The behaviour of cast & wrought AD730® is shown as a line for comparison. The grain size of the cast & wrought alloy is 10 in ASTM standard [29], which corresponds to a diameter of 11.2µm according to conversion tables.

For comparison, Figure 2 also shows the creep resistance of the cast & wrought AD730® after the same subsolvus and ageing heat-treatment (subsolvus solutionizing at 1080°C for 4 h + ageing at 760°C for 16 h). The creep resistance of the (B+C) alloy fabricated by laser powder bed fusion is similar to that of the cast & wrought alloy. The average grain diameter for the cast & wrought alloy is reported to be 11.2 µm according to ASTM standards [29]. The difference in grain size with the (B+C) alloy fabricated by additive manufacturing is not expected to have a significant effect on the creep resistance, as the AD730® is known to be quasi-insensitive to grain size below a testing temperature of 700°C [32].

Table 2: Times to rupture, in h, of the heat-treated samples under creep at 650°C and a load of 690 MPa (two samples per alloy).

| Alloy | Sample 1 | Sample 2 |
|---|---|---|
| (B+C+Zr) | 1311 | 161 |
| (B+C) | 2478 | 2150 |
| (C) | 1.1 | 0.2 |

*3.3. Microstructures in the (B+C+Zr) alloy*

In order to investigate how boron, carbon and zirconium contribute to hot cracking and creep resistance, the as-built state and heat-treated microstructures of the (B+C+Zr) alloy were investigated. Recall that this is the alloy exhibiting hot cracking in the as-built state and a moderate creep resistance in the heat-treated state.

Figure 3 shows the inverse pole figure (IPF) maps of the as-built and heat-treated microstructures. The build direction is denoted as BD, while X/Y stands for the second direction of the observation plane. The as-built microstructure is composed of γ phase grains of irregular shapes, with the <110> direction preferentially aligned with the build direction (Figure 3a). This kind of microstructure is commonly found in laser powder bed fusion microstructures of nickel-based superalloys [8,33] and austenitic stainless steels [34,35], and it is known to arise from epitaxial growth of grains on the side of the melt pool. The length per unit area of boundaries having a misorientation angle >15°, which is used here as a measure of the grain boundary density, is 129 mm/mm$^2$. No grain boundary was found to exhibit a twin relationship. On the IPF maps, the cracks are systematically located at high-angle grain boundaries (Figure 3c and d), as typically reported for hot cracking [36].

The γ grain structure in the heat-treated sample is similar to that in the as-built sample (Figure 3b). The grains have retained their irregular shape and their preferential orientation, while the grain boundary density has slightly decreased to 98 mm/mm$^2$. This decrease is likely caused by recovery effects and the disappearance of small grains during the heat-treatment (compare Figure 3a and b). From a practical point of view, the grain boundary density can be considered as fairly constant throughout the heat-treatment since a 25 % decrease is unlikely to drastically modify the solute distribution at grain boundaries.

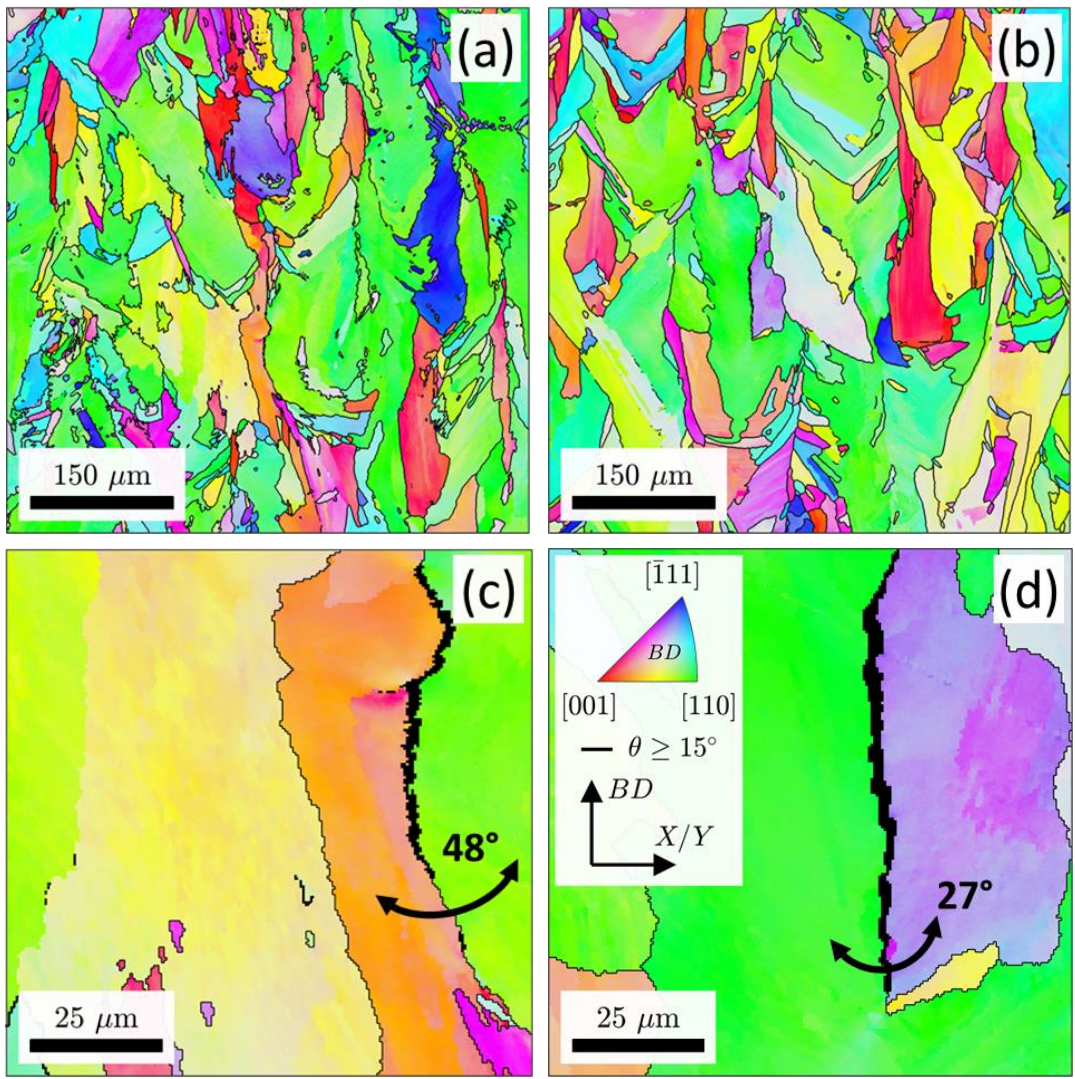

Figure 3 : BD-IPF maps of (a) the as-built microstructure, and (b) the heat-treated microstructure of the (B+C+Zr) alloy. (c) Enlarged view of a cracked grain boundary in the as-built microstructure. (d) Enlarged view of a cracked grain boundary in the heat-treated microstructure. The same colour code is applied for all maps.

While the γ grain structure remains practically unchanged, the precipitation state and the solute distribution are strongly affected by the heat-treatment. Figure 4 shows BSE-contrast images of the microstructure before and after the heat-treatment. In the as-built microstructure (Figure 4a), the grains inner structure contains columnar dendrites with limited secondary arms and sub-micrometric second-phase particles, which is typical of nickel-based superalloys processed by laser powder bed fusion [1,28,37–40].

The columnar dendritic spacing, revealed by compositional contrasts in the interdendritic regions, is about 0.7 µm. The particles, in bright and dark contrast, are preferentially located in the interdendritic regions. For more details, the interested reader is pointed to our recent work on the nature and formation sequence of the second-phase particles in the as-built microstructure [31].

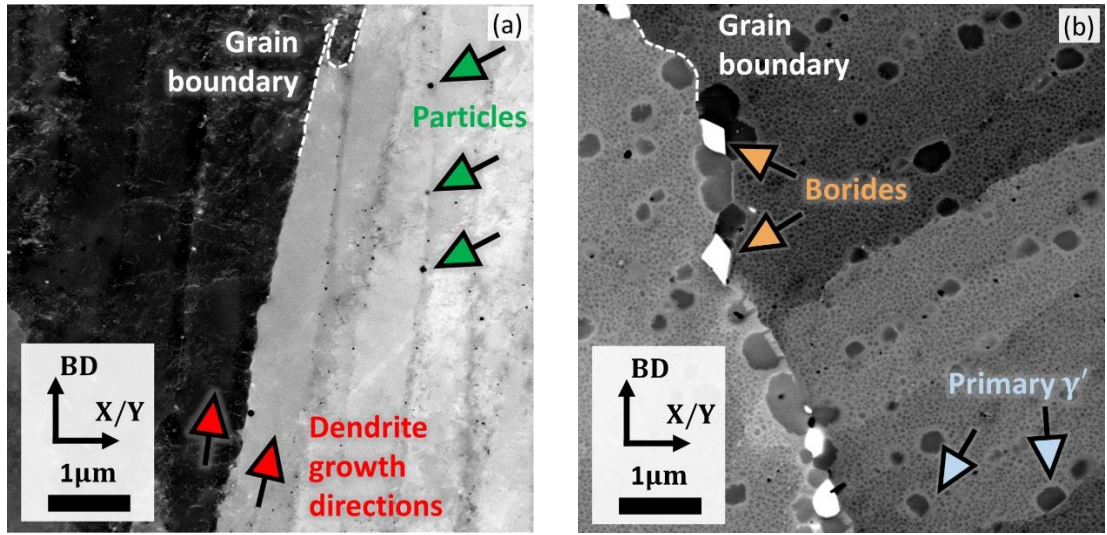

Figure 4: BSE contrast images of (a) the as-built microstructure, (b) the heat-treated microstructure.

In the heat-treated microstructure (Figure 4b), extensive precipitation of γ' particles is observed. Primary γ' particles, with a size of the order of 300-1000 nm, are found at grain boundaries and in the former interdendritic regions. The secondary γ' particles are approximately 50-100 nm in size and are distributed homogenously within the γ grains. The EDX maps in Figure 5a and the profiles in Figure 5b show that the bright-contrast particles of 300-1000 nm in size found at grain boundaries are chromium-molybdenum-tungsten borides. The chromium enrichment in borides is hardly visible due to the large fraction of chromium also present in the matrix (see Table 1). In Figure 4b and Figure 5a, dark-contrast particles of approximately 100 nm in size are also observed. By analogy with previous works [1,31], their peak of titanium suggests that they are carbonitrides, although in the present case the contrast with other phases is very faint. These particles are spread rather homogeneously in the microstructure as they can be found both at grain boundaries and within the γ grains.

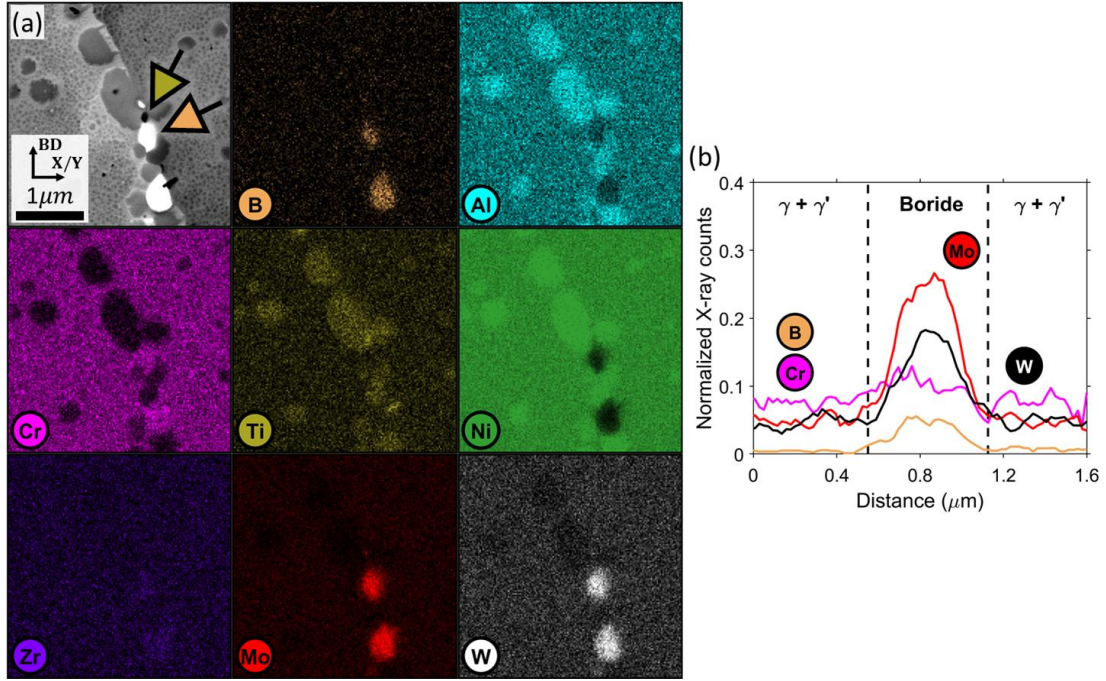

Figure 5: (a) BSE contrast image of the heat-treated microstructure near a grain boundary and corresponding X-ray maps. The colour contrast is proportional to the number of X-rays counted for each element. (b) X-ray count profile across a boride. The X-ray counts associated to each elements are normalized by the total number of counts at each position along the profile.

*3.4. Grain boundary segregation in the as-built microstructure of the (B+C+Zr) alloy*

Figure 6 shows a 3D atom probe reconstruction prepared from the uncracked region ahead of a partially cracked grain boundary of the as-built microstructure. The misorientation angle at the grain boundary is 41°, according to the preliminary EBSD observation shown in Figure 6a.

In Figure 6b, boron segregation along the grain boundary is shown with an iso-composition surface of 1.5 at.% B. The boron segregation extends on a rather large enrichment zone around the grain boundary, as illustrated by the iso-composition surface of 0.5 at.% B in Figure 6c. Figure 6d shows that nano-sized γ' precipitates are also located within this enrichment zone. Images taken from several viewing angles indicate that the precipitates have grown only on one side of the grain boundary. Their crystallographic structure was confirmed by TEM, as shown by Figure A. 1 of Appendix A. Finally, zirconium is found at the grain boundary and within the nano-sized γ' precipitates, as shown by the iso-composition surface of 0.3 at.% Zr in Figure 6e. The average composition of the γ' precipitates in zirconium was measured to be 0.6 at.%. Similar configurations were observed for other APT specimens prepared from the same grain boundary.

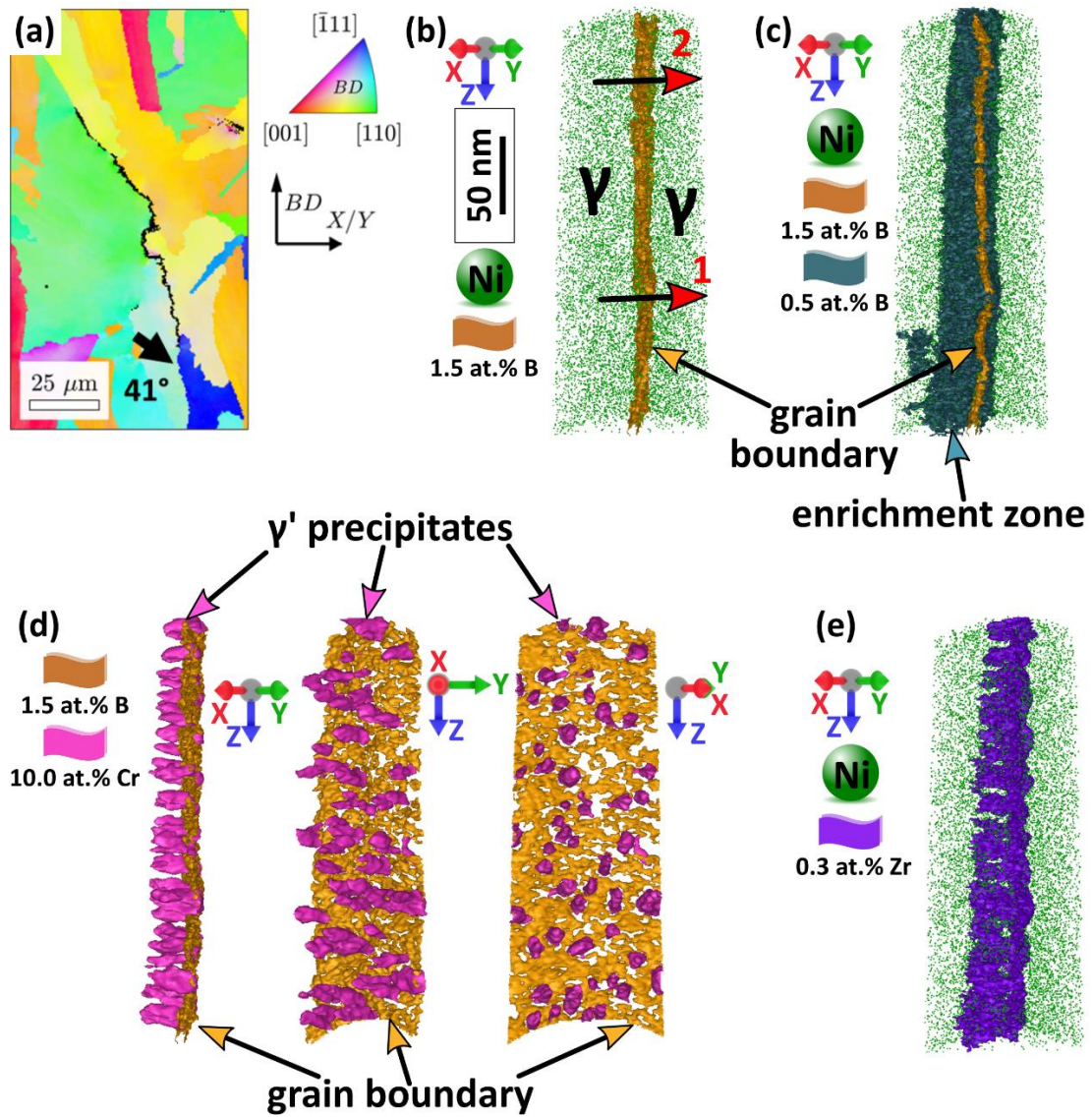

Figure 6: (a) BF-IPF map of the as-built microstructure showing in arrow the uncracked region of the partially cracked grain boundary selected for APT. (b) Atom probe reconstruction at the selected high-angle grain boundary, displayed with an iso-composition surface of 1.5 at.% B. (c) Same reconstruction showing an enrichment zone adjacent to the grain boundary by an iso-composition surface of 0.5 at.% B. (d) Atom probe reconstruction of the same specimen showing nano-sized γ' precipitates within the enrichment zone by an iso-composition surface of 10 at.% Cr, with views are shown taken from different angles. (e) An iso-composition surface of 0.3 at.% Zr showing segregation of Zr at the grain boundary and at the nano-sized γ' precipitates.

Figure 7 shows 1D composition profiles from cylindrical regions of interest (ROI) across the grain boundary denoted by arrow #1 and #2 in Figure 6b. The position and size of the cylindrical ROIs have been carefully selected to avoid including the γ' precipitates in the analysis. Segregation of boron, carbon and zirconium is clearly shown up to 2.0, 0.3 and 0.5 at.%, respectively (Figure 7a and b). In the enrichment zone adjacent to the grain boundary, segregation of the same elements is also observed, although at a lower level than at the grain boundary. With regard to the elements, molybdenum is found to segregate at the grain boundary, while titanium and aluminium are depleted, and tungsten remains at a constant concentration (Figure 7c and d). Chromium is slightly segregated at the grain boundary and in the enrichment zone, while nickel is depleted (Figure 7e and f).

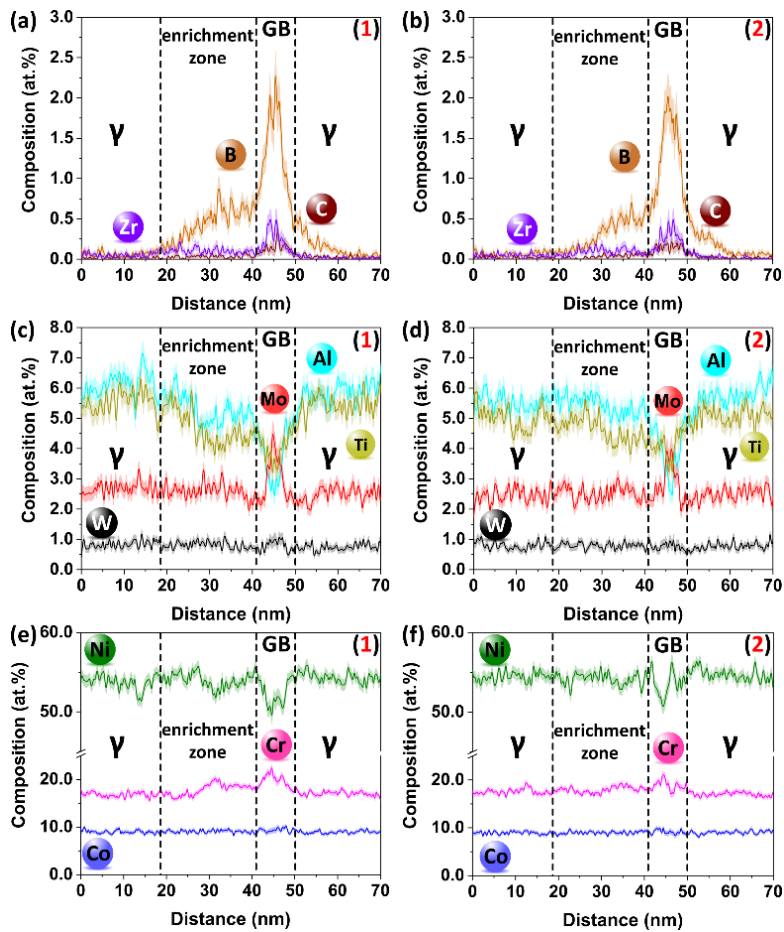

Figure 7: Two composition profiles across the grain boundary taken from the as-built microstructure. (a,b) boron, carbon and zirconium. (c,d) aluminium, titanium, molybdenum and tungsten. (e,f) chromium, cobalt and nickel. The number on each figure corresponds to the profile location denoted by the arrows in Figure 6b.

*3.5. Redistribution of solutes in the heat-treated microstructure of the (B+C+Zr) alloy*

Figure 8 shows a 3D atom probe reconstruction from the uncracked region ahead of partially cracked high angle grain boundary after the heat treatment. As shown in Figure 8a, the misorientation angle of the grain boundary is 45°, i.e. very similar to that investigated by APT in the as-built conditions (41°). The grain boundary in the APT reconstruction shown in Figure 8b is highlighted using an iso-composition surface of 1.5 at.% B. The secondary γ' precipitates are also visible thanks to their high content in aluminium.

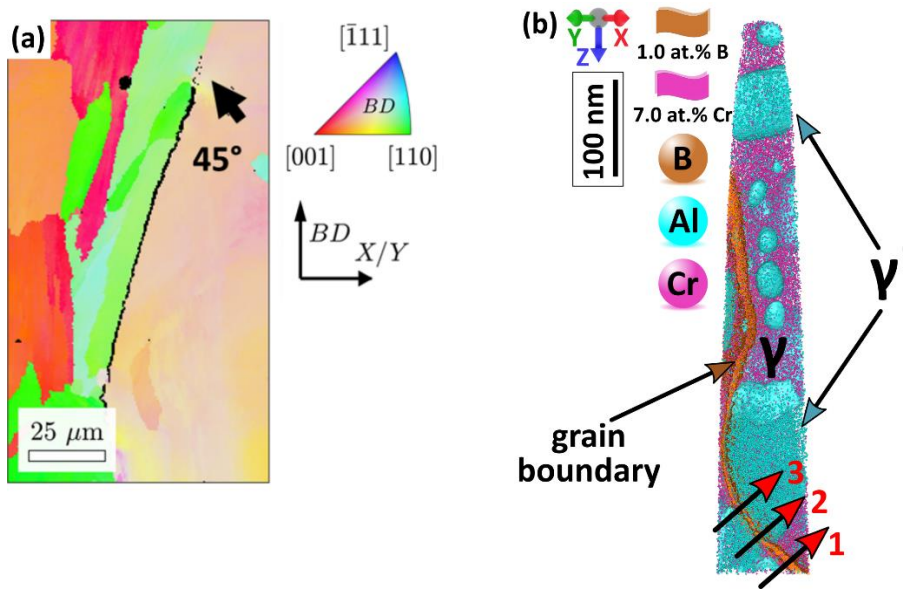

Figure 8: (a) BF-IPF map of the heat-treated microstructure showing in arrow the uncracked region of the partially cracked grain boundary selected for APT. (b) Atom probe reconstruction of a specimen including the grain boundary, displayed with 1.5 at.% and the γ' particles.

Figure 9 shows 1D composition profiles from cylindrical regions of interest (ROI) across the grain boundary as denoted by arrow #1, #2 and #3 in Figure 8b. The regions of interest were selected to cross the grain boundary at a γ/GB/γ, a γ/GB/γ', and a γ'/GB/ γ' interface. A significant difference compared to the as-built condition is the absence of zirconium grain boundary segregation (see Figure 9a-c). By contrast, grain boundary segregation of boron and carbon is clearly observed. The concentration of boron varies strongly as a function of the adjacent phases, being the highest when the grain boundary is between two grains of γ phase (Figure 9a - 6.0 at.%), and the lowest when the γ' phase is on both sides of the boundary (Figure 9c - 2.5 at.%). Segregation of molybdenum and tungsten is also found at the grain boundary (Figure 9d-f), and varies as a function of the nature of the phases on both sides of the boundary. Depletion of aluminium and titanium is found, as in the case of the as-fabricated sample. Segregation of chromium and cobalt can also be detected as a function of the nature of interface (Figure 9g-i).

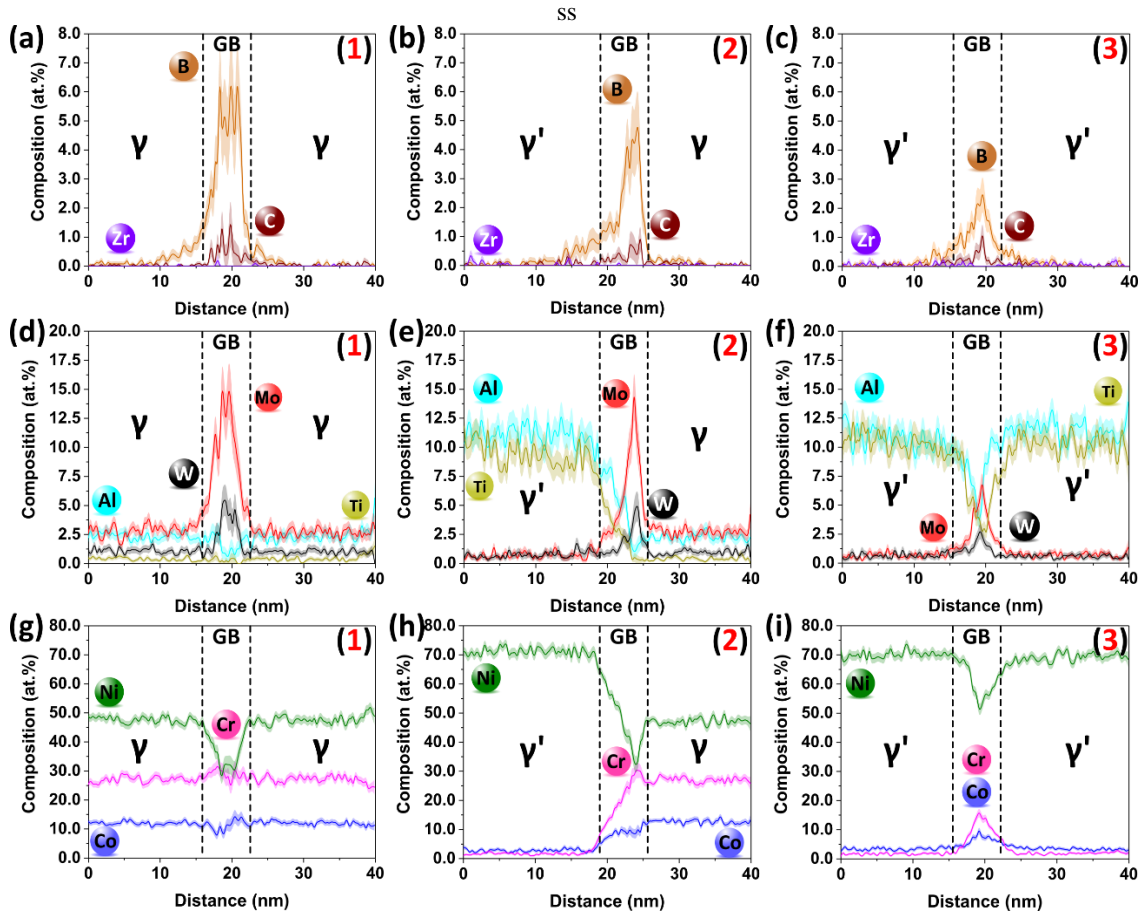

Figure 9: Composition profiles across the grain boundary taken from the heat-treated microstructure for a (γ- γ) interface, a (γ'- γ) interface and a (γ'- γ') interface. (a,b,c) boron, carbon and zirconium. (d,e,f) aluminium, titanium, molybdenum and tungsten. (g,h,i) chromium, nickel and cobalt. The number on each figure corresponds to the profile location denoted by arrows in Figure 8b.

*3.6. Presence of zirconium in γ' precipitates*

In an effort to locate where zirconium partitions during heat-treatment, the composition of the γ matrix phase and γ' precipitates in the heat-treated sample has been investigated. Figure 10 shows the zirconium composition profile across a γ/γ' interface away from the grain boundary in a heat-treated sample. Zirconium is found to partition preferentially within the γ' precipitates, which is consistent with another observation carried out on a different polycrystalline superalloy fabricated by electron powder bed fusion

[18].

It has been noticed earlier that zirconium also partitions preferentially within the nano-sized γ' precipitates adjacent to grain boundaries in the as-built microstructure (Figure 6e). Interestingly, there is a strong decrease in the zirconium composition of the γ' precipitates with heat-treatment. According to the atom probe reconstruction there is on average 0.6 at.% of zirconium in the nano-sized γ' precipitates of the as-built microstructure, but 0.04 at.% in the secondary γ' precipitates of the heat-treated microstructure. The measurements are summarized in Table 3. Variations of several at.% are also measured for the composition in other elements, such as molybdenum and tungsten .

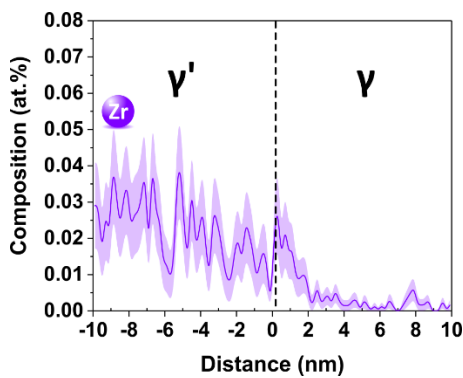

Figure 10: Zirconium composition profile across a γ/γ' interface away from the grain boundary.

Table 3: Composition in at.% of the γ and γ' particles in the as-built (As-b.) and heat-treated (HT) samples, extracted from atom probe reconstructions. For the as-built condition, the composition of the γ phase is taken far from the grain boundary, while that of the γ' precipitates is taken from the enrichment zone.

|  | Ni | Cr | Co | Mo | W | Al | Ti | Nb | C | B | Fe | Zr |
|---|---|---|---|---|---|---|---|---|---|---|---|---|
| γ in As-B. | 52.3 | 18.4 | 9.6 | 2.3 | 1.8 | 5.4 | 4.2 | 0.6 | 0.02 | 0.04 | 4.4 | 0.02 |
| γ' at GB in As-B. | 62.5 | 6.0 | 6.4 | 1.0 | 1.0 | 8.4 | 10.1 | 1.3 | 0.05 | 0.6 | 1.7 | 0.6 |
| γ in HT. | 48.4 | 26.2 | 11.6 | 2.6 | 1.1 | 2.0 | 0.9 | 0.2 | 0.02 | 0.02 | 6.0 | 0.01 |
| γ' in HT. | 70.8 | 11.9 | 3.0 | 0.5 | 0.7 | 11.1 | 10.0 | 1.3 | 0.01 | 0.02 | 0.7 | 0.04 |

## 4. Discussion

*4.1. Origin of the boron, carbon and zirconium segregation in the as-built microstructure*

Overall, the solute distribution in the as-built microstructure may be attributed to a combination of strong solute partitioning during solidification and limited diffusion in the solid state during solidification (i.e. back-diffusion), as well as limited diffusion in the solid state once solidification is completed (i.e. during cooling and reheating due to the melting of adjacent regions scanned by the laser). The limited diffusion during cooling may be assessed from the estimated diffusion distances given in Table 4 (see the data and formula used for calculating diffusion distances after isothermal and non-isothermal treatments in Appendix B). Assuming a cooling rate of $10^6 K.s^{-1}$ and an end-of-solidification temperature of 1000°C (corresponding to a solidified fraction of ~97% according to Scheil-Gulliver model predictions for the (B+C+Zr) alloy [31]), the estimated diffusion distance of boron, carbon and zirconium in the matrix are 390 nm, 370 nm and 44 nm, respectively, while the other major elements exhibit much lower diffusion distances. The predicted values are likely overestimated, as the selected diffusion coefficient do not account for the decrease of diffusivity with supersaturation (see e.g. [41]), and as the calculations do not account for the potential affinity of elements for grain boundaries and precipitates. As a result, even though boron, carbon and zirconium are usually considered as fast diffusers, the very short cooling time does not allow them to diffuse over distances larger than the interdendritic spacing.

Table 4: Estimated diffusion distances in meters. For fabrication, at linear decrease of the temperature from 1000°C to 20°C is assumed, with a cooling rate of $10^6$ K/s. The diffusion coefficients are taken from [1].

|  | B | C | Al | Ti | Cr | Zr | Mo | W |
|---|---|---|---|---|---|---|---|---|
| Fabrication | $3.9 \cdot 10^{-7}$ | $3.6 \cdot 10^{-7}$ | $1.5 \cdot 10^{-10}$ | $1.6 \cdot 10^{-10}$ | $5.1 \cdot 10^{-11}$ | $4.4 \cdot 10^{-8}$ | $4.7 \cdot 10^{-11}$ | $1.5 \cdot 10^{-11}$ |
| 4h at 1080°C | $8.8 \cdot 10^{-3}$ | $7.4 \cdot 10^{-3}$ | $5.9 \cdot 10^{-6}$ | $6.3 \cdot 10^{-6}$ | $2.1 \cdot 10^{-6}$ | $8.1 \cdot 10^{-4}$ | $1.8 \cdot 10^{-6}$ | $6.5 \cdot 10^{-7}$ |
| 16h at 760°C | $1.9 \cdot 10^{-3}$ | $2.0 \cdot 10^{-3}$ | $2.6 \cdot 10^{-7}$ | $2.7 \cdot 10^{-7}$ | $7.1 \cdot 10^{-8}$ | $2.6 \cdot 10^{-4}$ | $7.9 \cdot 10^{-8}$ | $2.1 \cdot 10^{-8}$ |

Thus, one can interpret grain boundary segregations in the as-built microstructure (Figure 7) as an evidence of the strong partitioning of boron, carbon and zirconium in the residual liquid during the last stages of solidification. This scenario can be drawn regardless of whether we assume the γ' precipitates to have formed by a eutectic transformation during solidification, as suggested by Wang et al. [28], or by a solid state transformation during cooling and reheating. In the case where γ' precipitates formed during eutectic transformation, the composition of the last liquid pockets is the same as that of the eutectic mixture, i.e. rich in boron, carbon and zirconium. In the case where they formed by a solid state transformation, the composition of the residual liquid in boron, carbon and zirconium is higher than that measured at the grain boundary by APT, and a fraction of these elements segregated at grain boundary will partition towards the γ' precipitates, thus contributing to the enrichment zone observed in Figure 7a and b.

*4.2. γ' precipitates and the enrichment zone at grain boundaries in the as-built microstructure*

It is noticeable that the estimated diffusion distances of boron, carbon and zirconium in the solid state are approximately 2 to 20 times larger than the width of the enrichment zone (20nm) shown in Figure 7a-b. This suggests that other mechanisms than mere diffusion may be involved in the formation of the enrichment zone. In that regard, it is interesting to recall that a strong correlation has been observed between the enrichment zone and the nano-sized γ' precipitates adjacent to the grain boundary.

The present observations are probably not sufficient to establish with certainty how nano-sized γ' precipitates form during fabrication. However, we believe that the asymmetric composition and precipitation profiles observed at the grain boundary may indicate that they have formed in the solid state. To support this argument, we point out that this configuration parallels, although at a finer scale, the fan-type morphologies of γ+γ' compounds found at grain boundaries of several polycrystalline nickel based superalloys cooled down from super-solvus heat-treatment at a rate below 3 $K.s^{-1}$ [42,43]. In addition, it can be noticed that the composition profiles of boron, carbon and zirconium measured at grain boundaries in the as-built microstructure (Figure 7a,b) follow similar trends as those predicted by Okamoto and Ågren [44] for the evolution of carbon behind a moving phase boundary affected by solute drag in steels. Following this point of view, the width of the enrichment zone shown in Figure 7a,b would represent the distance over which the grain boundary can migrate when this portion of the microstructure is in the heat-affected zone of the adjacent or stacked molten tracks. To summarize, the enrichment zone is suspected to result from γ' precipitation at grain boundaries migrating over short distances. Further work would be needed to fully assess this idea.

*4.3. Redistribution of boron, carbon and zirconium during heat-treatment.*

Although solid state diffusion is limited during fabrication, this is not the case during heat-treatment. The estimated length scale of the diffusion distance for boron, carbon and zirconium is of the order of $10^{-4}$-$10^{-3}$ m for both subsolvus annealing (4 h at 1080°C) and ageing (16 h at 760°C). The other elements are also estimated to diffuse over rather large distances, such as $10^{-7}$-$10^{-6}$ m for aluminium and $10^{-8}$-$10^{-7}$ m for tungsten (Table 4). It is only during ageing that aluminium, titanium, chromium, molybdenum and tungsten diffuse over distances lower than the interdendritic spacing. These estimations are consistent with the fact that the compositional contrast due to solidification microsegregation in the as-built microstructure (Figure 4a) is not observed in the heat-treated microstructure (Figure 4b). In comparison to the solute distribution in the as-built microstructure, that in the heat-treated sample can thus be considered as representative of thermodynamic equilibrium.

During heat-treatment, boron migrates mostly at grain boundaries. This is indicated by the formation of large borides at grain boundaries (Figure 4b and Figure 5) and by the strong increase in the boron content at grain boundaries (Figure 8 and Figure 9). It is reasonable to believe that these phenomena are driven by the migration of boron trapped in the interdendritic regions after fabrication towards the grain boundaries upon heat-treatment, since the interdendritic regions, being from a crystallographic point of view characterized as boundaries with no or very low misorientation angles, are not favourable sites for equilibrium segregation of boron [45]. Carbon exhibits a rather similar behaviour, since its concentration at grain boundaries increases with the heat-treatment. However, its affinity for grain boundaries may not be as high as boron, since some carbon may also be present in intragranular carbonitride particles.

The most striking behaviour is probably that of zirconium. In the as-built microstructure, zirconium is found at grain boundaries and in the γ' precipitates in the enrichment zone (Figure 6 and Figure 7). During heat treatment, zirconium diffuses away from grain boundaries (see Figure 9a-c) towards new γ' precipitates (Figure 10). Note that in textbooks on nickel-based superalloys, this element is systematically considered as segregating at grain boundaries in the heat-treated state [46,47]. This contradiction may seem surprising at a first sight, however it is also known that zirconium has a non-negligible solubility in γ' precipitates. Early reports indicate that up to 1.0 at.% of zirconium could partition in γ' precipitates [48], although such concentrations would result in a decomposition of the precipitates in other intermetallic particles at long annealing times [48]. The relative affinity of zirconium for γ' precipitates can likely be related to its electronic structure since other elements of the fourth column of the Mendeleev table, namely titanium and hafnium, also have a high solubility in γ' precipitates [48,49].

Per mole of material, the amount of zirconium contained within γ' precipitates and γ grains can be estimated by $[Zr] = f_{\gamma'} \times [Zr]_{\gamma'} + (1 - f_{\gamma'}) \times [Zr]_{\gamma}$, with $f_{\gamma'}$, the molar fraction of the γ' phase, and $[Zr]_{\gamma'}$, and $[Zr]_{\gamma}$, the molar fraction of zirconium in these two phases. After ageing at 760°C, the fraction of γ' is estimated to 38% according to thermodynamic equilibrium calculations [50]. The molar fraction of zirconium in γ' precipitates is 0.04 at.% while that in the γ phase is 0.01 at.% according to the stoichiometry of particles measured by APT (Table 3). The resulting amount of zirconium in γ' precipitates and γ grains is 0.0214 at.%, i.e. about 75% of the 0.0282 at.% of zirconium introduced initially in the nominal composition of the alloy (see Table 1). This is a crude estimate of the fraction of zirconium away from grain boundaries, and it does not account for its potential partitioning in second-phase particles (carbonitrides, oxides), but it does support the hypothesis that this element does not diffuse towards grain boundaries during heat-treatment.

*4.4. Designing crack-free and creep resistant superalloys for additive manufacturing*

As explained in the introduction, the mechanisms by which particular solutes strengthen grain boundaries during creep are complex and may vary as a function of the element. However, it seems universally admitted that, given their small fraction in the nominal composition of the alloy, these solutes must segregate at grain boundaries to have an effect on creep resistance [19,20,47]. In the present work, boron and carbon may be expected to act as grain boundary strengtheners as they segregate extensively at grain boundaries in the heat-treated microstructure (Figure 8 and Figure 9).

By contrast, zirconium is absent from grain boundaries in the heat-treated microstructure. Instead, the zirconium trapped at grain boundaries in the as-built microstructure migrates mostly towards γ' particles during the heat-treatment. In this configuration, one may expect zirconium to have no major and direct effect on the creep resistance of heat-treated components since 0.04 at.% of zirconium within γ' is *a priori* not sufficient to induce a significant change in their mechanical behaviour [48]. We do not expect that the defects generated during creep can induce a drastic change in the affinity of zirconium for grain boundaries.

With respect to hot cracking, the grain boundary segregation of boron, carbon and zirconium found in the as-built microstructure suggests that these elements may contribute to its occurrence during fabrication. It is known, in particular, that presence of both boron and zirconium is the combination of solutes most susceptible to lead to hot cracking [13,14]. Since zirconium is likely ineffective on creep resistance, it is suggested to remove it from the nominal composition of the alloy. This would allow the hot-cracking susceptibility of the alloy to be decreased without sacrificing the creep resistance. It would also provide more flexibility for the control of boron and carbon, which are more likely to play a role on creep resistance.

The potential benefit of removing zirconium from the nominal composition is supported by the variations of hot cracking susceptibility and creep resistance of the three alloys, as presented in Figure 1 and Figure 2. For example, the (B+C) alloy exhibits the best creep resistance, as good as the AD730® produced by the cast and wrought route, and is free of cracks after fabrication and heat treatment. On the other hand, the (B+C+Zr) alloy exhibits hot cracks after fabrication but its creep resistance is only moderately impacted by the cracks. We emphasize that this does not mean that zirconium is the sole cause of hot cracking. However, having achieved a crack-free and creep resistant component without zirconium is in itself a significant result given that this element is so often included in the nominal composition of polycrystalline superalloys for additive manufacturing [1–3,18].

By contrast, including boron to the nominal composition seems absolutely necessary to ensure an acceptable creep resistance, since the (C) alloy has an almost null resistance to creep. The extreme sensitivity of creep resistance to the content of boron is consistent with several previous works on nickel-based superalloys processed by conventional casting [20,21]. The boron content could thus likely be more refined than in the present study in order to maximize the creep lifetime while still avoiding hot cracking.

Two possible scenarios can explain the reason why zirconium has been usually considered as a grain boundary strengthener for polycrystalline superalloys, while the present results indicate the contrary. The first scenario would be that zirconium played a role on creep resistance at times where the impurity content of nickel-based superalloys, and especially the sulphur content, was not as well controlled as nowadays. For example, early works often assert that zirconium forms intermetallic particles with the sulphur located at grain boundaries, thus inhibiting the detrimental effect of this element on creep resistance [27]. The same studies have also shown that the effect of zirconium on creep resistance is not as significant when sulphur is intentionally maintained at negligible levels. In addition, more recent works using atom probe tomography tend to confirm that heat-treated microstructures of modern nickel-based superalloys produced by powder metallurgy [21] or additive manufacturing [17] do not exhibit equilibrium segregation of zirconium at grain boundaries. The second scenario would be that the extreme processing conditions of laser powder bed fusion have induced microstructures whose particular features (e.g. oxygen content, grain boundary structure) prevent equilibrium grain boundary segregation of zirconium from developing during heat-treatment. However, we find this second explanation less convincing, since the creep resistance of the alloy without zirconium is on par with that of the alloy processed by the cast & wrought route.

It can be remarked that zirconium is traditionally known to improve the resistance of nickel-based superalloys to air corrosion at high temperature [51,52]. Its addition is believed to improve the adherence of the protective alumina scale at the surface of the component, thus helping to increase its lifetime in service. In order to fully evaluate the viability of removing zirconium from the nominal composition of polycrystalline nickel-based superalloys for additive manufacturing, it would be interesting to investigate the consequences for corrosion resistance. One must keep in mind, however, that not all properties can be optimized at the same time, in particular for nickel-based superalloys submitted to extreme conditions of temperature, atmosphere and loading. In that regard, we expect that achieving a crack-free and creep resistant component will remain the most important aspect to consider for polycrystalline nickel-based superalloy processed by additive manufacturing.

## 5. Conclusions

Three versions of the AD730® polycrystalline superalloy with (B+C+Zr), (B+C) and (C) additions were processed by laser powder bed fusion to unravel the effect of those solutes on hot cracking sensitivity and creep lifetime. The (B+C+Zr) alloy exhibits significant hot cracking at high angle grain boundaries, while in the (B+C) and (C) alloys, hot cracking is suppressed. The absence of zirconium has no detrimental effect on the creep performance of the (B+C) alloy at 650°C. On the other hand, the absence of boron in alloy (C) results in a dramatic loss of creep lifetime.

The effect of these elements on hot-cracking and creep lifetime was then investigated at a more fundamental level in the (B+C+Zr) alloy with the help of grain boundary segregation measurements by atom probe tomography. The grain boundary segregation of boron, carbon and zirconium at grain boundaries were studied in the as-built conditions as well as after heat treatment.

It was shown that if included in the nominal composition of the alloy, boron, carbon and zirconium segregate at high-angle grain boundaries in the as-built microstructure as a result of solute partitioning and limited solid-state diffusion during fabrication. Nano-sized $\gamma'$ precipitates were found to be distributed at only one side of the grain boundary. Their presence seem correlated with an enrichment zone with intermediate levels of segregation. The non-symmetric composition and precipitation profiles at grain boundaries in the as-built microstructure is possibly related to grain boundary migration allowed during the cycled heat treatments induced when melting stacked or adjacent tracks.

After heat-treatment, there is no equilibrium segregation of zirconium at grain boundaries, as zirconium migrates towards newly formed $\gamma'$ precipitates. This suggests that zirconium has no effect on the strength of grain boundaries during creep. By contrast, equilibrium grain boundary segregation of boron and carbon further increases, and is found to depend on the combination of the adjacent phases, i.e. $\gamma/GB/\gamma$, $\gamma/GB/\gamma'$ and $\gamma'/GB/\gamma'$.

These observations allow us to suggest guidelines for crack-free and creep resistant superalloys produced by additive manufacturing. In particular, the presence of zirconium in the nominal composition does not seem justified as it promotes hot-cracking and seems to have no significant effect on the creep lifetime. Its removal may give more flexibility to control the elements that truly play a role on creep, such as boron and to a lesser extent carbon.

**Acknowledgements**

This work was financially supported by Aubert & Duval as part of the SOFIA project (BPI France) and has benefited from the characterization equipment of the Grenoble INP - CMTC platform supported by the Centre of Excellence of Multifunctional Architectured Materials "CEMAM" n°ANR-10-LABX-44-01 funded by the Investments for the Future programme. S.A. would like to acknowledge financial support from the Alexander Humbolt foundation. Batman is also gratefully acknowledged for fruitful discussions. Uwe Tezins and Andreas Sturm are acknowledged for their support to the FIB & APT facilities at MPIE.

**Appendix A. TEM imaging of γ' precipitates at grain boundaries**

Thin foils of the as-built microstructure of the (B+C+Zr) alloy were prepared by jet electro-polishing in a solution of 5% perchloric acid in methanol. Observations were conducted using a Jeol 3010 equipped with a LaB6 electron source operating at 200kV. Figure A. 1 shows a dark-field image of the as-built microstructure near a grain boundary containing γ' precipitates with $L1_2$ ordering. Visually, the fraction of those precipitates appears much lower than in the work of Wang on a CM247LC alloy [28].

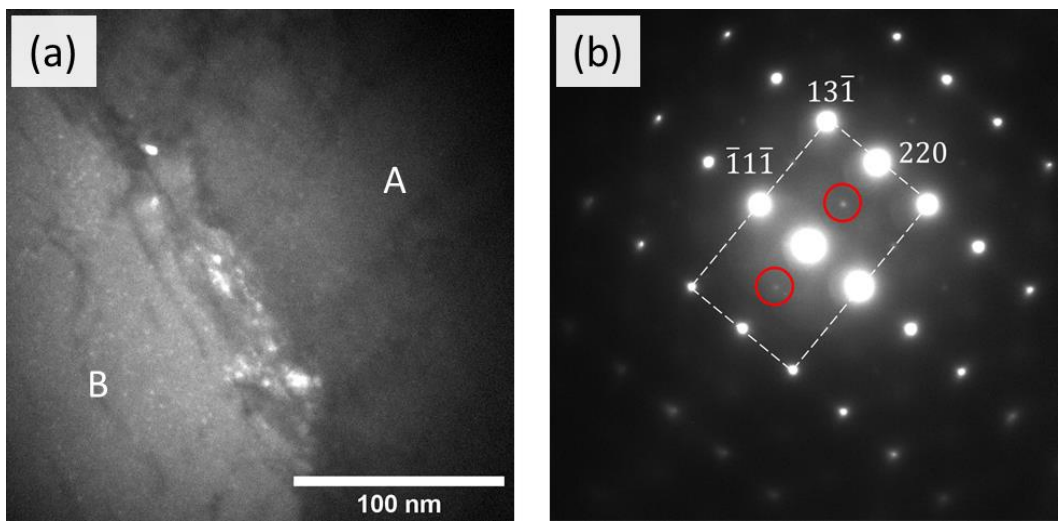

Figure A. 1: (a) dark-field image from γ' reflections at a grain boundary in the as-built microstructure. (b) [$1\bar{1}\bar{2}$] zone axis diffraction pattern acquired in grain A. The red circles denote the 110 reflections associated with the γ' phase.

**Appendix B. Diffusion distances for isothermal and non-isothermal treatments**

During isothermal treatment, atoms migrate away from their original position by a random walk mechanism. We assume migration to take place in one dimension to represent a front of atoms diffusing away from a grain boundary or an interdendritic region. After a time $t$, the displacement $x$ of the atoms relative to their origin follows a Gaussian distribution centred on 0, with quadratic mean:

$$<x^2> = 2Dt$$

With $D$ the diffusivity of the atoms in the matrix.

In this context, the mean diffusion distance corresponds to the mean of the displacement norms. If the displacements follow a Gaussian distribution, their norms must follow a half-normal distribution, whose mean is proportional to the quadratic mean of the distribution of positions:

$$<|x|> = <\sqrt{x^2}> = \sqrt{\frac{2}{\pi}}\sqrt{<x^2>}$$

During anisothermal treatment, the diffusivities vary as a function of time (or equivalently, temperature). Thus, the quadratic mean of the displacement is given by:

$$<x^2> = 2\int_0^t D(t)dt$$

Or, setting K as the cooling rate:

$$<x^2> = \frac{2}{K}\int_{T_{solidus}}^{T_{ambiant}} D(T)dT$$

This integral is solved numerically since it has no real analytical solution. The numerical integration converges to a stable value when choosing temperature increments below 1°C. The mean diffusion distance is then calculated using the same formula as for the isothermal treatment, i.e. $<|x|> = \sqrt{\frac{2}{\pi}}\sqrt{<x^2>}$.

The diffusivities are assumed to follow an Arrhenius law $D = D_0 \exp(-Q/RT)$, with $D_0$ the pre-exponential factor, $Q$ the activation energy, $R$ the universal gas constant and $T$ the temperature. The parameters taken for the simulations are given in Table A. 1. They have been taken from a previous work on additive manufacturing of nickel-based superalloys ref. [1]. The rather low activation energy of zirconium compared to other heavy elements is related to its high compressibility in the nickel matrix [53].

Table A. 1: diffusion coefficients of elements in nickel, according to Hariharan et al. [1].

| Parameter | B | C | Al | Ti | Cr | Zr | Mo | W |
|---|---|---|---|---|---|---|---|---|
| $D_0$ (m$^2$/s) | $3.27 \cdot 10^{-4}$ | $3.70 \cdot 10^{-5}$ | $8.79 \cdot 10^{-5}$ | $1.07 \cdot 10^{-4}$ | $9.25 \cdot 10^{-5}$ | $4.60 \cdot 10^{-6}$ | $1.09 \cdot 10^{-5}$ | $6.90 \cdot 10^{-6}$ |
| $Q$ (kJ/mol) | 163.5 | 147.0 | 276.0 | 276.9 | 299.0 | 132.1 | 278.8 | 297.1 |

**References**

[1] A. Hariharan, L. Lu, J. Risse, A. Kostka, B. Gault, E.A. Jägle, D. Raabe, Misorientation-dependent solute enrichment at interfaces and its contribution to defect formation mechanisms during laser additive manufacturing of superalloys, Phys. Rev. Mater. 3 (2019) 123602. https://doi.org/10.1103/PhysRevMaterials.3.123602.

[2] Y.T. Tang, C. Panwisawas, J.N. Ghoussoub, Y. Gong, J.W.G. Clark, A.A.N. Németh, D.G. McCartney, R.C. Reed, Alloys-by-design: Application to new superalloys for additive manufacturing, Acta Mater. 202 (2021) 417–436. https://doi.org/10.1016/j.actamat.2020.09.023.

[3] S. Griffiths, H. Ghasemi Tabasi, T. Ivas, X. Maeder, A. De Luca, K. Zweiacker, R. Wróbel, J. Jhabvala, R.E. Logé, C. Leinenbach, Combining alloy and process modification for micro-crack mitigation in an additively manufactured Ni-base superalloy, Addit. Manuf. 36 (2020) 101443. https://doi.org/10.1016/j.addma.2020.101443.

[4] S. Griffiths, H. Ghasemi-Tabasi, A. De Luca, J. Pado, S.S. Joglekar, J. Jhabvala, R.E. Logé, C. Leinenbach, Influence of Hf on the heat treatment response of additively manufactured Ni-base superalloy CM247LC, Mater. Charact. 171 (2021) 110815. https://doi.org/10.1016/j.matchar.2020.110815.

[5] N.J. Harrison, I. Todd, K. Mumtaz, Reduction of micro-cracking in nickel superalloys processed by Selective Laser Melting: A fundamental alloy design approach, Acta Mater. 94 (2015) 59–68. https://doi.org/10.1016/j.actamat.2015.04.035.

[6] E. Chauvet, P. Kontis, E.A. Jägle, B. Gault, D. Raabe, C. Tassin, J.-J. Blandin, R. Dendievel, B. Vayre, S. Abed, G. Martin, Hot cracking mechanism affecting a non-weldable Ni-based superalloy produced by selective electron Beam Melting, Acta Mater. 142 (2018) 82–94. https://doi.org/10.1016/j.actamat.2017.09.047.

[7] L.N. Carter, X. Wang, N. Read, R. Khan, M. Aristizabal, K. Essa, M.M. Attallah, Process optimisation of selective laser melting using energy density model for nickel based superalloys, Mater. Sci. Technol. 32 (2016) 657–661. https://doi.org/10.1179/1743284715Y.0000000108.

[8] C. Qiu, H. Chen, Q. Liu, S. Yue, H. Wang, On the solidification behaviour and cracking origin of a nickel-based superalloy during selective laser melting, Mater. Charact. 148 (2019) 330–344. https://doi.org/10.1016/j.matchar.2018.12.032.

[9] Y. Chen, F. Lu, K. Zhang, P. Nie, S.R. Elmi Hosseini, K. Feng, Z. Li, Dendritic microstructure and hot cracking of laser additive manufactured Inconel 718 under improved base cooling, J. Alloys Compd. 670 (2016) 312–321. https://doi.org/10.1016/j.jallcom.2016.01.250.

[10] A. Ramakrishnan, G.P. Dinda, Direct laser metal deposition of Inconel 738, Mater. Sci. Eng. A. 740–741 (2019) 1–13. https://doi.org/10.1016/j.msea.2018.10.020.

[11] M. Rappaz, J.-M. Drezet, M. Gremaud, A new hot-tearing criterion, Metall. Mater. Trans. A. 30 (1999) 449–455. https://doi.org/10.1007/s11661-999-0334-z.

[12] N. Wang, S. Mokadem, M. Rappaz, W. Kurz, Solidification cracking of superalloy single- and bi-crystals, Acta Mater. 52 (2004) 3173–3182. https://doi.org/10.1016/j.actamat.2004.03.047.

[13] J. Grodzki, N. Hartmann, R. Rettig, E. Affeldt, R.F. Singer, Effect of B, Zr, and C on Hot Tearing of a Directionally Solidified Nickel-Based Superalloy, Metall. Mater. Trans. A. 47 (2016) 2914–2926. https://doi.org/10.1007/s11661-016-3416-8.

[14] J. Zhang, R.F. Singer, Effect of Zr and B on castability of Ni-based superalloy IN792, Metall. Mater. Trans. A. 35 (2004) 1337–1342. https://doi.org/10.1007/s11661-004-0308-0.

[15] W.F. Savage, Effect of Minor Elements on Hot- Cracking Tendencies of Inconel 600, (n.d.) 9.

[16] S. Antonov, D. Isheim, D.N. Seidman, S. Tin, Phosphorous behavior and its effect on secondary phase formation in high refractory content powder-processed Ni-based superalloys, Materialia. 7 (2019) 100423. https://doi.org/10.1016/j.mtla.2019.100423.

[17] Y. Zhou, A. Volek, Effect of carbon additions on hot tearing of a second generation nickel-base superalloy, Mater. Sci. Eng. A. 479 (2008) 324–332. https://doi.org/10.1016/j.msea.2007.06.076.

[18] P. Kontis, E. Chauvet, Z. Peng, J. He, A.K. da Silva, D. Raabe, C. Tassin, J.-J. Blandin, S. Abed, R.

Dendievel, B. Gault, G. Martin, Atomic-scale grain boundary engineering to overcome hot-cracking in additively-manufactured superalloys, Acta Mater. 177 (2019) 209–221. https://doi.org/10.1016/j.actamat.2019.07.041.

[19] G.W. Meetham, Trace elements in superalloys–an overview, Met. Technol. 11 (1984) 414–418. https://doi.org/10.1179/030716984803275188.

[20] D.M. Shah, D.N. Duhl, Effect of Minor Elements on the Deformation Behavior of Nickel-Base Superalloys, in: Superalloys 1988 Sixth Int. Symp., TMS, 1988: pp. 693–702. https://doi.org/10.7449/1988/Superalloys_1988_693_702.

[21] P. Kontis, H.A.M. Yusof, S. Pedrazzini, M. Danaie, K.L. Moore, P.A.J. Bagot, M.P. Moody, C.R.M. Grovenor, R.C. Reed, On the effect of boron on grain boundary character in a new polycrystalline superalloy, Acta Mater. 103 (2016) 688–699. https://doi.org/10.1016/j.actamat.2015.10.006.

[22] D. Blavette, P. Duval, L. Letellier, M. Guttmann, Atomic-scale APFIM and TEM investigation of grain boundary microchemistry in Astroloy nickel base superalloys, Acta Mater. 44 (1996) 4995–5005. https://doi.org/10.1016/S1359-6454(96)00087-0.

[23] P.A.J. Bagot, O.B.W. Silk, J.O. Douglas, S. Pedrazzini, D.J. Crudden, T.L. Martin, M.C. Hardy, M.P. Moody, R.C. Reed, An Atom Probe Tomography study of site preference and partitioning in a nickel-based superalloy, Acta Mater. 125 (2017) 156–165. https://doi.org/10.1016/j.actamat.2016.11.053.

[24] H. Zhang, O.A. Ojo, M. Chaturvedi, Nanosize boride particles in heat-treated nickel base superalloys, 58 (2008) 167–170.

[25] T. Alam, P.J. Felfer, M. Chaturvedi, L.T. Stephenson, M.R. Kilburn, J.M. Cairney, Segregation of B, P, and C in the Ni-Based Superalloy, Inconel 718, Metall. Mater. Trans. A. 43 (2012) 2183–2191. https://doi.org/10.1007/s11661-012-1085-9.

[26] Y.T. Tang, A.J. Wilkinson, R.C. Reed, Grain Boundary Serration in Nickel-Based Superalloy Inconel 600: Generation and Effects on Mechanical Behavior, Metall. Mater. Trans. A. 49 (2018) 4324–4342. https://doi.org/10.1007/s11661-018-4671-7.

[27] M. McLean, A. Strang, Effects of trace elements on mechanical properties of superalloys, Met. Technol. 11 (1984) 454–464. https://doi.org/10.1179/030716984803274800.

[28] X. Wang, L.N. Carter, B. Pang, M.M. Attallah, M.H. Loretto, Microstructure and yield strength of SLM-fabricated CM247LC Ni-Superalloy, Acta Mater. 128 (2017) 87–95. https://doi.org/10.1016/j.actamat.2017.02.007.

[29] Brochure AD730®, (n.d.). https://www.aubertduval.com/wp-media/uploads/2017/05/2017_Brochure_AD730.pdf.

[30] S. Sulzer, Z. Li, S. Zaefferer, S.M. Hafez Haghighat, A. Wilkinson, D. Raabe, R. Reed, On the assessment of creep damage evolution in nickel-based superalloys through correlative HR-EBSD and cECCI studies, Acta Mater. 185 (2020) 13–27. https://doi.org/10.1016/j.actamat.2019.07.018.

[31] A. Després, C. Mayer, M. Veron, E.F. Rauch, M. Bugnet, J.-J. Blandin, G. Renou, C. Tassin, P. Donnadieu, G. Martin, On the variety and formation sequence of second-phase particles in nickel-based superalloys fabricated by laser powder bed fusion, Materialia. 15 (2021) 101037. https://doi.org/10.1016/j.mtla.2021.101037.

[32] L. Thébaud, P. Villechaise, J. Cormier, C. Crozet, A. Devaux, D. Béchet, J.-M. Franchet, A. Organista, F. Hamon, Relationships between Microstructural Parameters and Time-Dependent Mechanical Properties of a New Nickel-Based Superalloy AD730$^{TM}$, Metals. 5 (2015) 2236–2251. https://doi.org/10.3390/met5042236.

[33] D. Grange, J.D. Bartout, B. Macquaire, C. Colin, Processing a non-weldable nickel-base superalloy by Selective Laser Melting: role of the shape and size of the melt pools on solidification cracking, Materialia. 12 (2020) 100686. https://doi.org/10.1016/j.mtla.2020.100686.

[34] O. Andreau, I. Koutiri, P. Peyre, J.-D. Penot, N. Saintier, E. Pessard, T. De Terris, C. Dupuy, T. Baudin, Texture control of 316L parts by modulation of the melt pool morphology in selective laser melting, J. Mater. Process. Technol. 264 (2019) 21–31. https://doi.org/10.1016/j.jmatprotec.2018.08.049.

[35] M.-S. Pham, B. Dovgyy, P.A. Hooper, C.M. Gourlay, A. Piglione, The role of side-branching in

microstructure development in laser powder-bed fusion, Nat. Commun. 11 (2020) 749. https://doi.org/10.1038/s41467-020-14453-3.
[36] J.C. Lippold, Welding Metallurgy and Weldability, (n.d.) 421.
[37] P. Kanagarajah, F. Brenne, T. Niendorf, H.J. Maier, Inconel 939 processed by selective laser melting: Effect of microstructure and temperature on the mechanical properties under static and cyclic loading, Mater. Sci. Eng. A. 588 (2013) 188–195. https://doi.org/10.1016/j.msea.2013.09.025.
[38] I. Lopez-Galilea, B. Ruttert, J. He, T. Hammerschmidt, R. Drautz, B. Gault, W. Theisen, Additive manufacturing of CMSX-4 Ni-base superalloy by selective laser melting: Influence of processing parameters and heat treatment, Addit. Manuf. 30 (2019) 100874. https://doi.org/10.1016/j.addma.2019.100874.
[39] O.M.D.M. Messé, R. Muñoz-Moreno, T. Illston, S. Baker, H.J. Stone, Metastable carbides and their impact on recrystallisation in IN738LC processed by selective laser melting, Addit. Manuf. 22 (2018) 394–404. https://doi.org/10.1016/j.addma.2018.05.030.
[40] R. Muñoz-Moreno, V.D. Divya, S.L. Driver, O.M.D.M. Messé, T. Illston, S. Baker, M.A. Carpenter, H.J. Stone, Effect of heat treatment on the microstructure, texture and elastic anisotropy of the nickel-based superalloy CM247LC processed by selective laser melting, Mater. Sci. Eng. A. 674 (2016) 529–539. https://doi.org/10.1016/j.msea.2016.06.075.
[41] B. Lawrence, C.W. Sinclair, M. Perez, Carbon diffusion in supersaturated ferrite: a comparison of mean-field and atomistic predictions, Model. Simul. Mater. Sci. Eng. 22 (2014) 065003. https://doi.org/10.1088/0965-0393/22/6/065003.
[42] V.V. Atrazhev, S.F. Burlatsky, D.V. Dmitriev, D. Furrer, N.Y. Kuzminyh, I.L. Lomaev, D.L. Novikov, S. Stolz, P. Reynolds, The Mechanism of Grain Boundary Serration and Fan-Type Structure Formation in Ni-Based Superalloys, Metall. Mater. Trans. A. 51 (2020) 3648–3657. https://doi.org/10.1007/s11661-020-05790-5.
[43] H.L. Danflou, M. Macià, T. Sanders, T. Khan, Mechanisms of Formation of Serrated Grain Boundaries in Nickel Base Superalloys, (1996). https://doi.org/10.7449/1996/SUPERALLOYS_1996_119_127.
[44] R. Okamoto, J. Ågren, A model for interphase precipitation based on finite interface solute drag theory, Acta Mater. 58 (2010) 4791–4803. https://doi.org/10.1016/j.actamat.2010.05.016.
[45] G. Miyamoto, A. Goto, N. Takayama, T. Furuhara, Three-dimensional atom probe analysis of boron segregation at austenite grain boundary in a low carbon steel - Effects of boundary misorientation and quenching temperature, Scr. Mater. 154 (2018) 168–171. https://doi.org/10.1016/j.scriptamat.2018.05.046.
[46] R.C. Reed, The Superalloys: Fundamentals and Applications, (n.d.) 390.
[47] M.J. Donachie, S.J. Donachie, Superalloys: a technical guide, 2. ed., 3. print, ASM, Materials Park, Ohio, 2008.
[48] J.E. Doherty, B.H. Kea, A.F. Giamei, On the origin of the ductility enhancement in Hf-doped Mar-M200, JOM. 23 (1971) 59–62. https://doi.org/10.1007/BF03355744.
[49] P.S. Kotval, J.D. Venables, R.W. Calder, The role of hafnium in modifying the microstructure of cast nickel-base superalloys, Metall. Trans. (n.d.) 6.
[50] A. Devaux, B. Picqué, M.F. Gervais, E. Georges, T. Poulain, P. Héritier, AD730™ - A New Nickel-Based Superalloy for High Temperature Engine Rotative Parts, in: Superalloys 2012, John Wiley & Sons, Ltd, 2012: pp. 911–919. https://doi.org/10.1002/9781118516430.ch100.
[51] A.S. Kahn, C.E. Lowell, C.A. Barrett, The Effect of Zirconium on the Isothermal Oxidation of Nominal Ni - 14Cr - 24Al Alloys, J. Electrochem. Soc. 127 (1980) 670. https://doi.org/10.1149/1.2129731.
[52] C.A. Barrett, A.S. Khan, C.E. Lowell, The Effect of Zirconium on the Cyclic Oxidation of NiCrAl Alloys, J. Electrochem. Soc. 128 (1981) 25. https://doi.org/10.1149/1.2127382.
[53] C.Z. Hargather, S.-L. Shang, Z.-K. Liu, A comprehensive first-principles study of solute elements in dilute Ni alloys: Diffusion coefficients and their implications to tailor creep rate, Acta Mater. 157 (2018) 126–141. https://doi.org/10.1016/j.actamat.2018.07.020.

# Supplementary materials

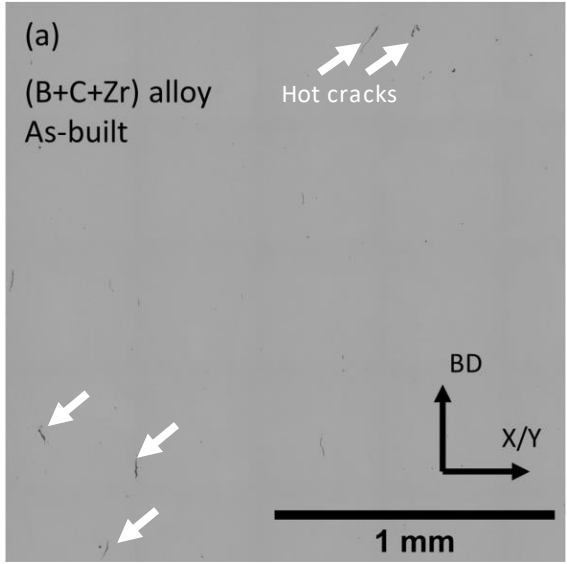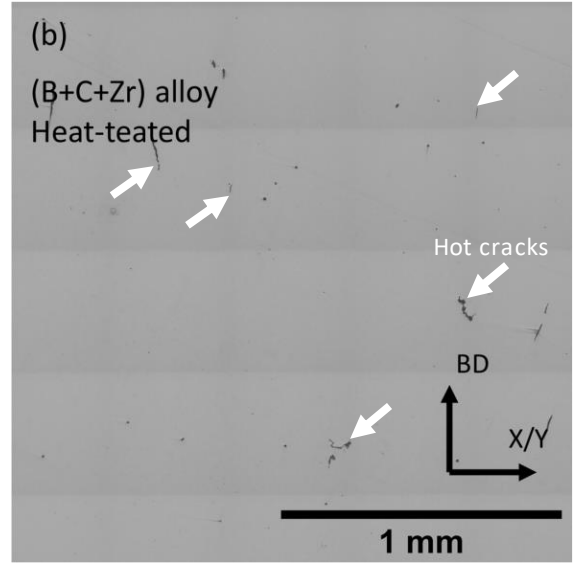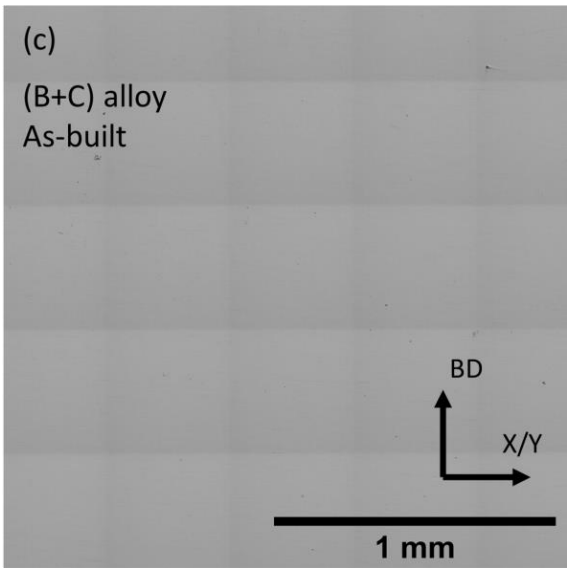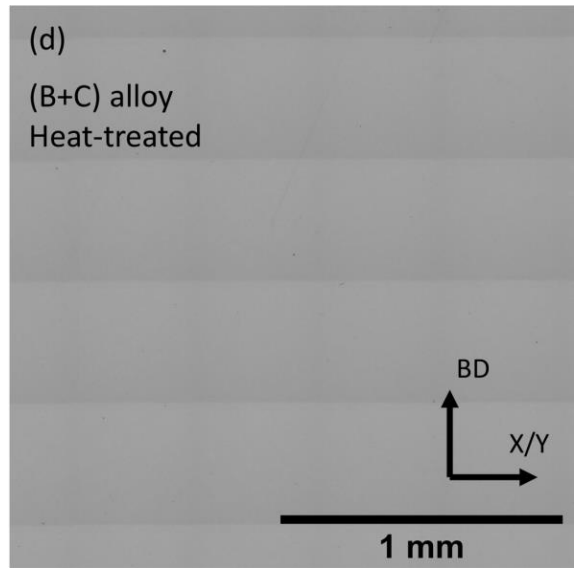

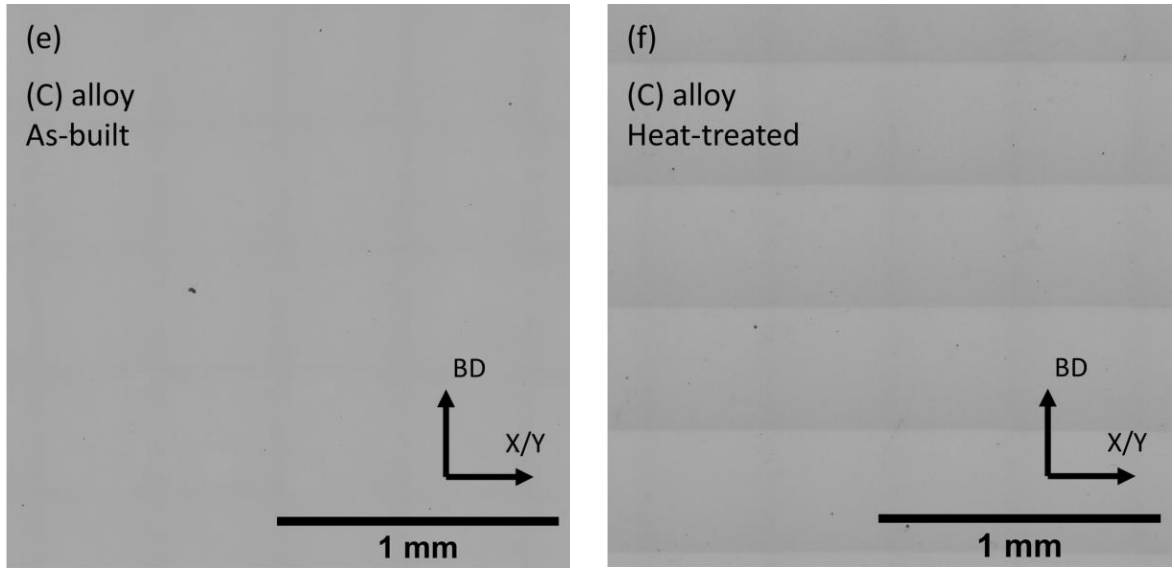

Figure S. 1: optical micrographs of (a-b) the as-built and heat-treated microstructures of the (B+C+Zr) alloy respectively, (c-d) the as-built and heat-treated microstructures of the (B+C) alloy respectively, (e-f) the as-built and heat-treated microstructures of the (B+C+Zr) alloy respectively.

Table S. 1: Summary of the areas investigated and measured crack densities.

| Alloy and condition | Area investigated (mm$^2$) | Crack density (mm/mm$^2$) |
|---|---|---|
| (B+C+Zr) as-built | 12.8 | 0.124±0.03 |
| (B+C+Zr) heat-treated | 29 | 0.123±0.03 |
| (B+C) as-built | 23.9 | 0 |
| (B+C) heat-treated | 29.4 | 0 |
| (C) as-built | 20.2 | 0 |
| (C) heat-treated | 29.4 | 0 |